\documentclass[superscriptaddress,preprint,amsmath,amssymb,aps,pre]{revtex4-2}

\usepackage{graphicx}
\usepackage{amsmath}
\usepackage{dcolumn}
\usepackage{txfonts}
\usepackage{bm}
\usepackage{mathrsfs}
\usepackage{amsfonts}
\usepackage{amssymb}
\usepackage{appendix}
\usepackage{braket}
\usepackage{color}
\usepackage[colorlinks=true,citecolor=magenta,anchorcolor=blue]{hyperref}

\begin{document}

\title{Extended-localized transition in diffusive quasicrystals}

\author{Zhoufei Liu}
\affiliation{Department of Physics, State Key Laboratory of Surface Physics, and Key Laboratory of Micro and Nano Photonic Structures (MOE), Fudan University, Shanghai 200438, China}

\author{Pei-Chao Cao}
\author{Ying Li}\email{eleying@zju.edu.cn}
\affiliation{State Key Laboratory of Extreme Photonics and Instrumentation, ZJU-Hangzhou Global Scientific and Technological Innovation Center, Zhejiang University, Hangzhou 310027, China}
\affiliation{International Joint Innovation Center, Key Laboratory of Advanced Micro/Nano Electronic Devices $\&$ Smart Systems of Zhejiang, The Electromagnetics Academy of Zhejiang University, Zhejiang University, Haining 314400, China}
\affiliation{Shaoxing Institute of Zhejiang University, Zhejiang University, Shaoxing 312000, China}

\author{Jiping Huang}\email{jphuang@fudan.edu.cn}
\affiliation{Department of Physics, State Key Laboratory of Surface Physics, and Key Laboratory of Micro and Nano Photonic Structures (MOE), Fudan University, Shanghai 200438, China}

\date{\today}

\begin{abstract}

Compared to periodic systems, quasicrystals without translational invariance exhibit unexpected localization properties. The extended-localized transition in quasicrystals has been observed in both quantum and classical wave systems. However, its manifestation in diffusion systems, which serve as novel platforms for exploring phases of matter in condensed matter physics, remains unexplored. Here, we present the implementation of the extended-localized transition in a diffusive quasicrystal based on the coupled ring chain structure. By modulating the thermal conductivities of rings, we obtain the diffusive one-dimensional Aubry-Andr${\rm{\acute{e}}}$-Harper (AAH) model, which exhibits an extended-localized transition. Thanks to the ring-shaped chain, we clearly demonstrate the extended-localized transition under the uniform excitation through temperature field simulations. For the localized state, the temperature field clearly demonstrates a multiple localization centers phenomenon, which has no counterpart in wave systems. We also quantitatively investigate the temperature evolution and size effect of this transition. Furthermore, the local excitation has been adopted to demonstrate the temperature field for both the extended and localized states. Besides, we implement the non-Hermitian diffusive AAH model by rotating rings, whose temperature field shows a moving multiple localization centers phenomenon in the localized phase. Finally, we give the experimental suggestions for the diffusive AAH model and propose a potential application named as double-trace distributed generator. Our results can facilitate the design of flexible thermal devices and efficient heat management. 

\end{abstract}

\maketitle

\section{\label{sec:introduction}Introduction}

Anderson localization is an ancient but everlasting research topic in condensed matter physics~\cite{AndersonPR58}. A common belief is that any infinitesimal disorder leads to the localization of all eigenstates in one-dimensional systems without the extended-localized transition~\cite{AbrahamsPRL79, LeeRMP85}. However, it has been discovered that the extended-localized transition can occur in quasiperiodic systems at a finite transition point, such as the Aubry-Andr${\rm{\acute{e}}}$-Harper (AAH) model~\cite{Aubry80, Harper55}. The prototypical AAH model has been experimentally realized in both quantum~\cite{RoatiNat08, WangPRL22, LinPRL22} and classical wave~\cite{LahiniPRL09, VerbinPRL13, ZhangPRL23, WangPRB23} systems. Besides, there has been a great deal of interest in studying the extended-localized transition of the non-Hermitian AAH model recently~\cite{ZengPRA17, LonghiPRL19, LonghiPRB19, JiangPRB19, LiuPRB21-1, LiuPRB21-2}, which presents new opportunities for quasicrystals. 

Diffusion systems, inherently dissipative in nature, play a critical role in heat and mass transfer. Thermal metamaterials~\cite{YangPR21, LiNRM21, ZhangNRP23, YangRMP24, JuAM22} enable diverse and flexible manipulation of heat flow, including cloaking~\cite{FanAPL08, ChenAPL08, GaoEPL13, WangJAP18, JinPNAS23}, illusion~\cite{ZhuAIP15, XuJAP18, HuAM18, HuAM19, YangESEE19, JinRes23}, and chameleonlike behaviours~\cite{XuPRAP19, YangPRAP20}. Recently, topological phases of matter and non-Hermitian physics have been realized in diffusion systems~\cite{Liu24}, such as exceptional point~\cite{LiSci19, XuPRL21, CaoSA24}, Weyl exceptional ring~\cite{XuPNAS22}, robust edge state~\cite{XuNP22}, one-dimensional Su-Schrieffer-Heeger model~\cite{YoshidaSR21, QiAM22, HuAM22}, higher-order topological insulator~\cite{LiuPRL24, XuNC23, WuAM23}, and non-Hermitian skin effect~\cite{CaoCP21, CaoCPL22, HuangCPL23, LiuSB24}. Diffusion systems are widely regarded as excellent platforms for exploring novel phases in condensed matter physics. 

However, the quasiperiodic extended-localized transition has yet to be observed in diffusion systems. The primary challenge lies in the non-local evolution for excitations commonly used in naturally dissipative diffusion systems. While previous studies have worked on diffusive localized states (i.e. topological edge modes)~\cite{XuNP22, QiAM22, HuAM22}, these modes decay rapidly without a clear observation in the temperature field. Moreover, the observation of these localized states necessitates specific initial conditions to align with their decay rates, further complicating their studies. Thus, uncovering diffusive localized states represents a highly inconvenient yet crucial task.

In this paper, we propose an extended-localized transition based on the diffusive quasicrystal. The diffusive AAH model is constructed by the coupled ring chain structure. By modulating the thermal conductivities of rings, we obtain the diffusive counterpart of AAH model, which exhibits both the diffusive extended and localized states. Through temperature field simulations, we clearly illustrate the extended-localized transition under the uniform excitation. Specifically, the temperature field corresponding to the diffusive localized state exhibits a phenomenon of multiple localization centers obviously, which is unique in diffusion systems without the wave counterpart. A quantitative analysis on this transition is performed by investigating the temperature evolution and size effect. Besides, we change the initial condition as the local excitation to show both the extended and localized states. Furthermore, we introduce the anti-parity-time reversal (APT) symmetry into diffusive AAH model. This non-Hermitian model demonstrates a moving multiple localization centers phenomenon in the localized phase. Finally, we provide the experimental suggestions for the diffusive AAH model and present a potential application dubbed as double-trace distributed generator. Our work offers new insights into efficient and robust heat manipulation and provides a distinct mechanism of heat insulation.

The outline for the rest of this paper is as follows. In Sec.~\ref{sec:coupled_ring}, we introduce the coupled ring chain structure and derive its effective Hamiltonian. In Sec.~\ref{sec:diffusive_AAH}, we get the diffusive AAH model by mapping the original AAH model onto coupled ring chain structure and discuss its properties. In Sec.~\ref{sec:uniform_excit}, we perform the temperature field simulation of the extended-localized transition under the uniform excitation. We also quantitatively investigate the temperature evolution and size effect of this transition. In Sec.~\ref{sec:local_excit}, we discuss the extended-localized transition under the local excitation. In Sec.~\ref{sec:APT}, we implement the diffusive APT symmetric AAH model and perform the temperature field simulation. We give the experimental suggestions in Sec.~\ref{sec:exper_sugg} and propose the potential application in Sec.~\ref{sec:potent_appli}. Finally, our conclusion is given in Sec.~\ref{sec:conclusion}.

\section{\label{sec:coupled_ring}Coupled ring chain structure}

\begin{figure}
\includegraphics[width=0.5\linewidth]{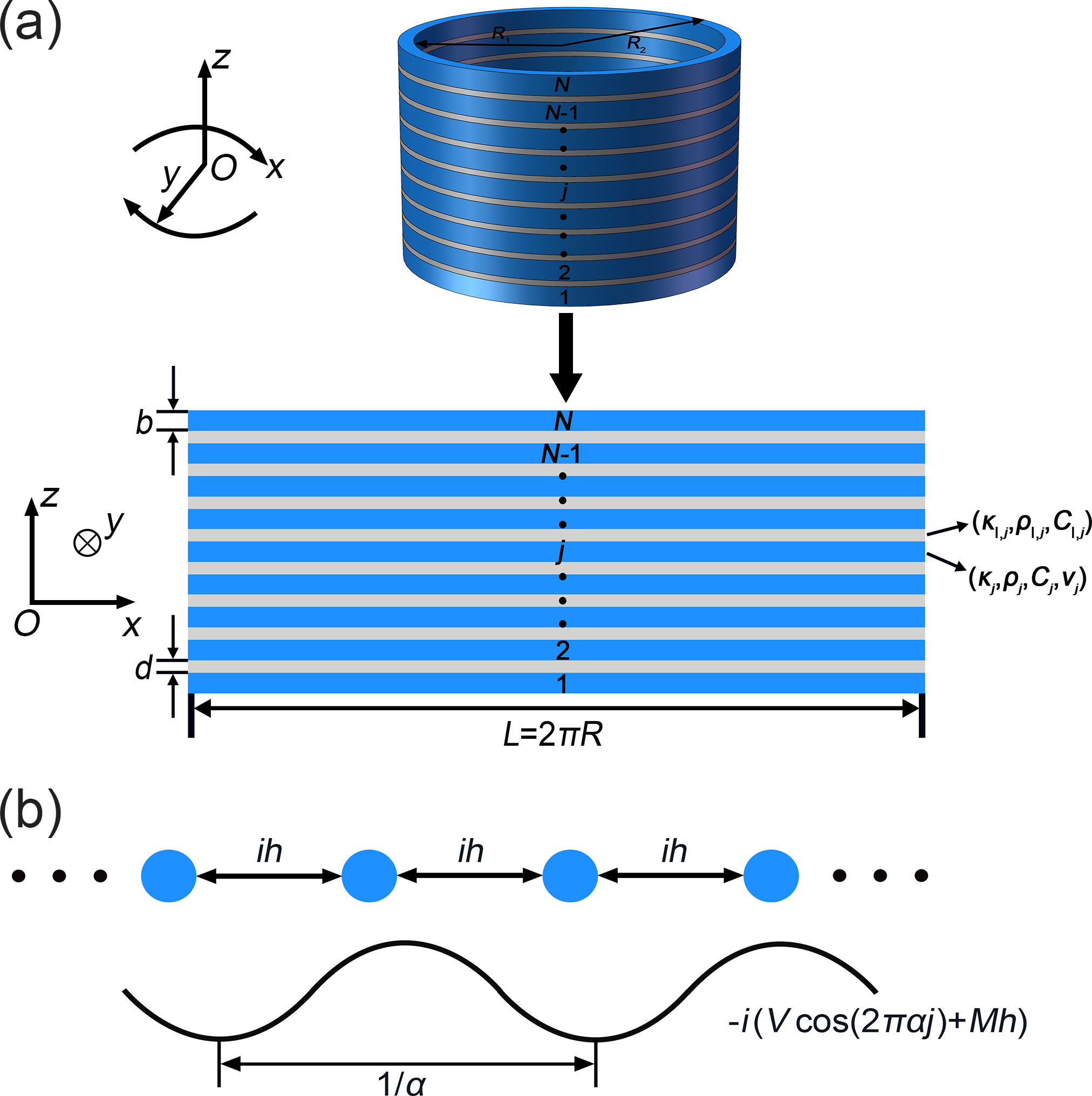}
\caption{Schematic diagram of diffusive AAH model. (a) Coupled ring chain structure. The blue area denotes the ring, while the grey area represents the interlayer. The diagram is presented in 3D for clarity. To simplify the depiction, we have transformed the rings into planar channels and set the periodic boundary condition at both ends, as shown in the lower diagram. Here, the bottom channel is labelled as the first ring, while the top channel is denoted as the $N$-th ring. (b) Equivalent tight-binding model, with the onsite potential $-i\left(V{\rm{cos}}(2{\pi}{\alpha}j)+Mh\right)$.}
\label{Fig1}
\end{figure}

We start by considering the coupled ring chain structure, as depicted in Fig.~\ref{Fig1}(a). The structure consists of several rings vertically coupled in the $z$ direction to form a chain through interlayers. We denote the interior and exterior radii of the rings by $R_{1}$ and $R_{2}$, respectively. For simplicity, we assume that $R_{1} \approx R_{2} \approx R$ and that the ring's perimeter is $L = 2\pi R$. The thickness of the ring and interlayer are denoted by $b$ and $d$, respectively. Using Fourier's law of heat conduction, we can write the thermal coupling equation for the $j$-th ring as
\begin{align}\label{1}
&\frac{\partial T_{j}(x,t)}{\partial t}=\frac{\kappa_j}{\rho_j C_j}\frac{\partial^2 T_{j}(x,t)}{\partial x^2}+v_j\frac{\partial T_{j}(x,t)}{\partial x}\nonumber \\
&+h_{j-1,j}\left[T_{j-1}(x,t)-T_{j}(x,t)\right]+h_{j,j}\left[T_{j+1}(x,t)-T_{j}(x,t)\right],
\end{align}  
where $T_{j}(x,t)$, $v_j$, $\kappa_j$, $\rho_j$, and $C_j$ are the temperature field, rotating velocity, thermal conductivity, mass density, and heat capacity of the $j$-th ring, respectively, and $x$ is the position along the ring. The heat exchange rate between the $(j-1)$-th interlayer and the $j$-th ring is denoted by $h_{j-1,j}=\kappa_{\mathrm{I},j-1}/(\rho_j C_j b d)$, where $\kappa_{\mathrm{I},j-1}$ is the thermal conductivity of the $(j-1)$-th interlayer. Similarly, the heat exchange rate between the $j$-th interlayer and the $j$-th ring is denoted by $h_{j,j}=\kappa_{\mathrm{I},j}/(\rho_j C_j b d)$. As the temperature field of the ring is periodic, we can assume that Eq.~\ref{1} has a plane wave solution of the form $T_{j}(x,t)=A_{j}e^{i(\beta x-\omega t)}$, where $A_j$ is the amplitude of the temperature field of the $j$-th ring, and $\omega$ is the decay rate. Here, $\beta=2m\pi/L=m/R$ is the propagation constant, where $m$ is the mode order. In this study, we focus on the fundamental mode ($m=1$) because only the slowest decaying mode can be clearly observed in diffusion systems. Substituting the plane wave solution into Eq.~\ref{1}, we obtain the effective Hamiltonian of the coupled ring chain structure under the open boundary condition. The Hamiltonian can be written in the second-quantized form:
\begin{equation}\label{2}
\hat{H}=i\sum_{j}\left[h_{j,j+1}\hat{c}_{j+1}^{\dagger}\hat{c}_{j}+h_{j,j}\hat{c}_{j}^{\dagger}\hat{c}_{j+1}+(S_{j}+i{\beta}v_{j})\hat{c}_{j}^{\dagger}\hat{c}_{j}\right],
\end{equation}
where $i=\sqrt{-1}$ denotes the imaginary unit, $\hat{c}_{j}^{\dagger}$ and $\hat{c}_{j}$ are the creation and annihilation operators of the $j$-th ring, respectively. The rotation term can be introduced to investigate the non-Hermitian effect in thermal diffusion, which corresponds to the gain and loss in wave systems~\cite{LiSci19}. The onsite term $S_{j}$ takes the form of $S_{j}=-\left({\beta}^{2}D_{j}+h_{j-1,j}+h_{j,j}\right)$ in the bulk and $S_{1(N)}=-\left({\beta}^{2}D_{1(N)}+h_{1,1(N-1,N)}\right)$ at the boundary. Here, $D_{j}=\kappa_{j}/(\rho_{j}C_{j})$ represents the diffusivity of the $j$-th ring. Furthermore, $N$ denotes the total number of rings. This ring-shaped chain can help localize the temperature field, and thus contribute to the implementation of diffusive localized state.

\section{\label{sec:diffusive_AAH}Diffusive AAH model}

\begin{figure}
\includegraphics[width=0.6\linewidth]{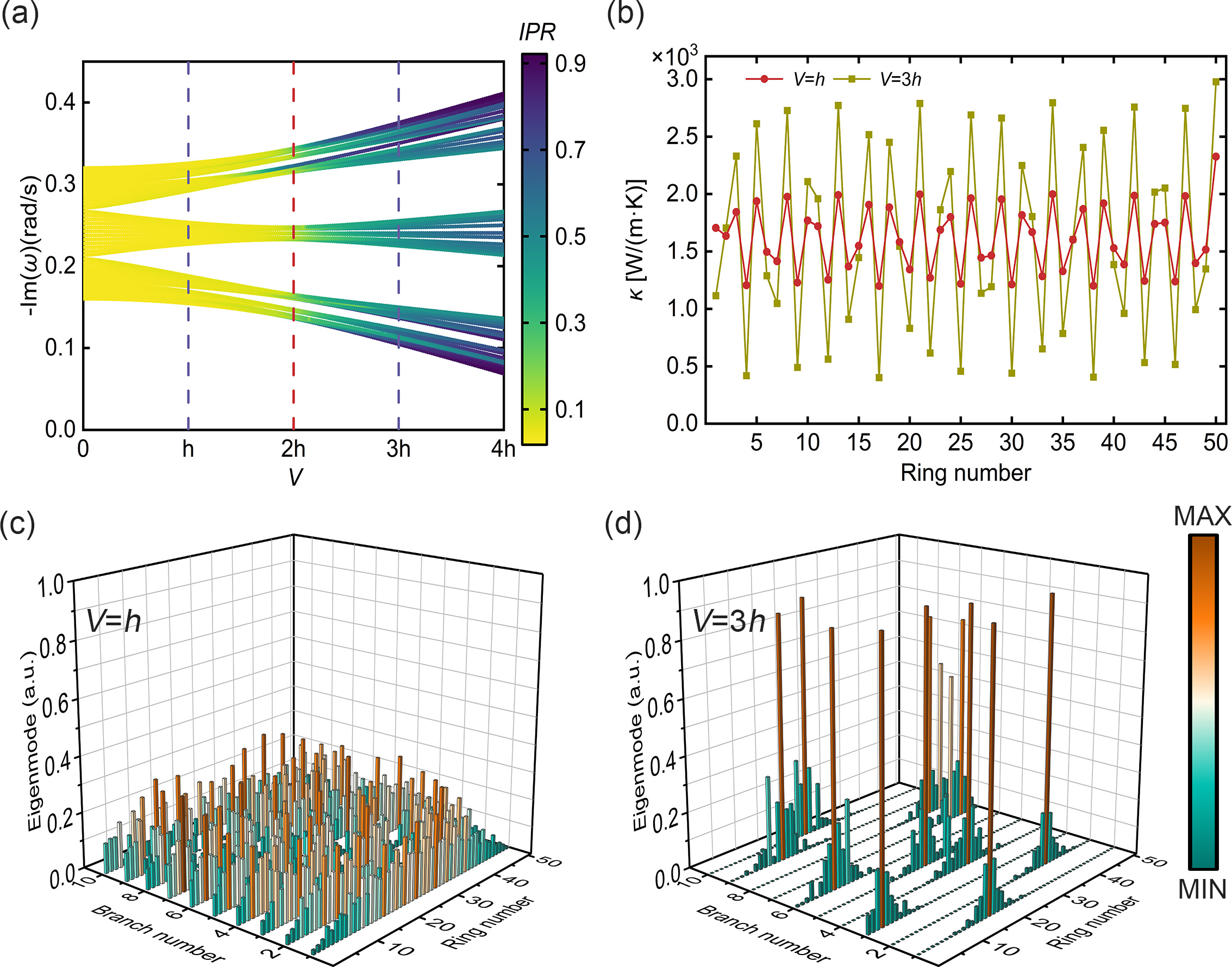}
\caption{Diffusive AAH model. (a) Decay rates with different strengths of quasiperiodic potential. The colorbar indicates the $IPR$. The red dashed line marks the extended-localized transition point $V=2h$. (b) The adjusted thermal conductivities of each ring for the extended state ($V=h$) and localized state ($V=3h$), which are marked as purple dashed lines in Fig.~\ref{Fig2}(a). Eigenstate distributions of the ten slowest decaying branches for the (c) extended state and (d) localized state. The parameters are $b=12.5$~mm, $d=2$~mm, $R=100$~mm, $\rho=1000$~kg/m$^3$, $C=1000$~J/(kg$\cdot$K), $\kappa_{\rm{I}}=1$~W/(m$\cdot$K), $M=6$, and $\alpha=(\sqrt{5}-1)/2$. The number of rings is $N=50$.}
\label{Fig2}
\end{figure}

Then we aim to obtain the anti-Hermitian diffusive AAH Hamiltonian by mapping the Hermitian AAH model onto the ring chain's Hamiltonian. The original AAH Hamiltonian is given by
\begin{equation}\label{3} 
\hat{H}=\sum_{j}[t\hat{c}_{j+1}^{\dagger}\hat{c}_{j}+t\hat{c}_{j}^{\dagger}\hat{c}_{j+1}+V{\rm{cos}}(2{\pi}{\alpha}j)\hat{c}_{j}^{\dagger}\hat{c}_{j}]
\end{equation}
where $t$ is the hopping amplitude, $V$ is the strength of the quasiperiodic potential, and $\alpha$ is an irrational number, usually chosen as the inverse golden ratio. To obtain the anti-Hermitian diffusive AAH Hamiltonian, we multiply the original Hamiltonian by $i$ and replace $t$ with the heat exchange rate $h$. Additionally, we should add a constant term $Mh$ to the onsite potential to ensure that the adjusted thermal conductivities of rings are positive, where $M$ is a constant. The effective diffusive AAH Hamiltonian can be written as
\begin{equation}\label{4}
\hat{H}=i\sum_{j}\left[h\hat{c}_{j+1}^{\dagger}\hat{c}_{j}+h\hat{c}_{j}^{\dagger}\hat{c}_{j+1}-\left(V{\rm{cos}}(2\pi\alpha{j})+Mh\right)\hat{c}_{j}^{\dagger}\hat{c}_{j}\right]
\end{equation}
where $h=\kappa_{\rm{I}}/(\rho Cbd)$ is the heat exchange rate. The equivalent tight-binding model is demonstrated in Fig.~\ref{Fig1}(b). The decay rates with different strengths of the quasiperiodic potential are shown in Fig.~\ref{Fig2}(a). Here, we use the inverse participation ratio ($IPR$) to characterize the degree of localization for each eigenstate~\cite{LonghiPRL19, LiuPRB21-1, LiuPRB21-2}. The $IPR$ is defined as $IPR=\sum_{j}|\psi_{j}(E)|^4/(\sum_{j}|\psi_{j}(E)|^2)^2$, where $\psi_{j}(E)$ is the $j$-th component of the eigenstate corresponding to energy $E$. For an extended state, $IPR{\approx}1/N$ and approaches zero as $N{\rightarrow}\infty$. On the other hand, $IPR{\approx}1$ for a fully localized state. A clear extended-localized transition can be observed for each branch in the spectrum. The eigenstate distributions of the ten slowest decaying branches for the extended state ($V=h$) and the localized state ($V=3h$) are shown in Figs.~\ref{Fig2}(c,d). For the extended state, the eigenstates are almost evenly distributed. While for the localized state, each eigenstate is localized at the certain ring. Furthermore, the Lyapunov exponent is also an important quantity to characterize the localization. We have extracted the Lyapunov exponent of diffusive AAH model from the theoretical and simulated eigenstate distributions, which is discussed in the Appendix~\ref{app:AppA}. An obvious extended-localized transition can also be demonstrated in the Lyapunov exponent (see Fig.~\ref{Fig8}).

To meet the requirements of the AAH model, it is necessary to adjust the parameters of rings and interlayers. The velocities of rings should be set to zero because the diffusive AAH model is anti-Hermitian, which corresponds to the Hermitian AAH model in wave systems. The thermal conductivity of the ring is the only parameter that requires adjustment. By comparing the diffusive AAH Hamiltonian with the ring chain's Hamiltonian, the following correspondence can be observed:
\begin{equation}\label{5} 
\begin{aligned}
S_{1}=-(\beta^2D_{1}+h)&=-\left(V{\rm{cos}}(2\pi\alpha)+Mh\right), \\ 
S_{j}=-(\beta^2D_{j}+2h)&=-\left(V{\rm{cos}}(2\pi{\alpha}j)+Mh\right), \\ 
S_{N}=-(\beta^2D_{N}+h)&=-\left(V{\rm{cos}}(2\pi{\alpha}N)+Mh\right), 
\end{aligned}
\end{equation}
where $j=2,\cdots,N-1$. By solving these equations, the adjusted thermal conductivities of the rings can be obtained. Here we present two examples of adjusted thermal conductivities for the extended state ($V=h$) and localized state ($V=3h$) in Fig.~\ref{Fig2}(b).

\section{\label{sec:uniform_excit}Temperature field simulations of diffusive AAH model under the uniform excitation} 

\begin{figure}
\includegraphics[width=0.75\linewidth]{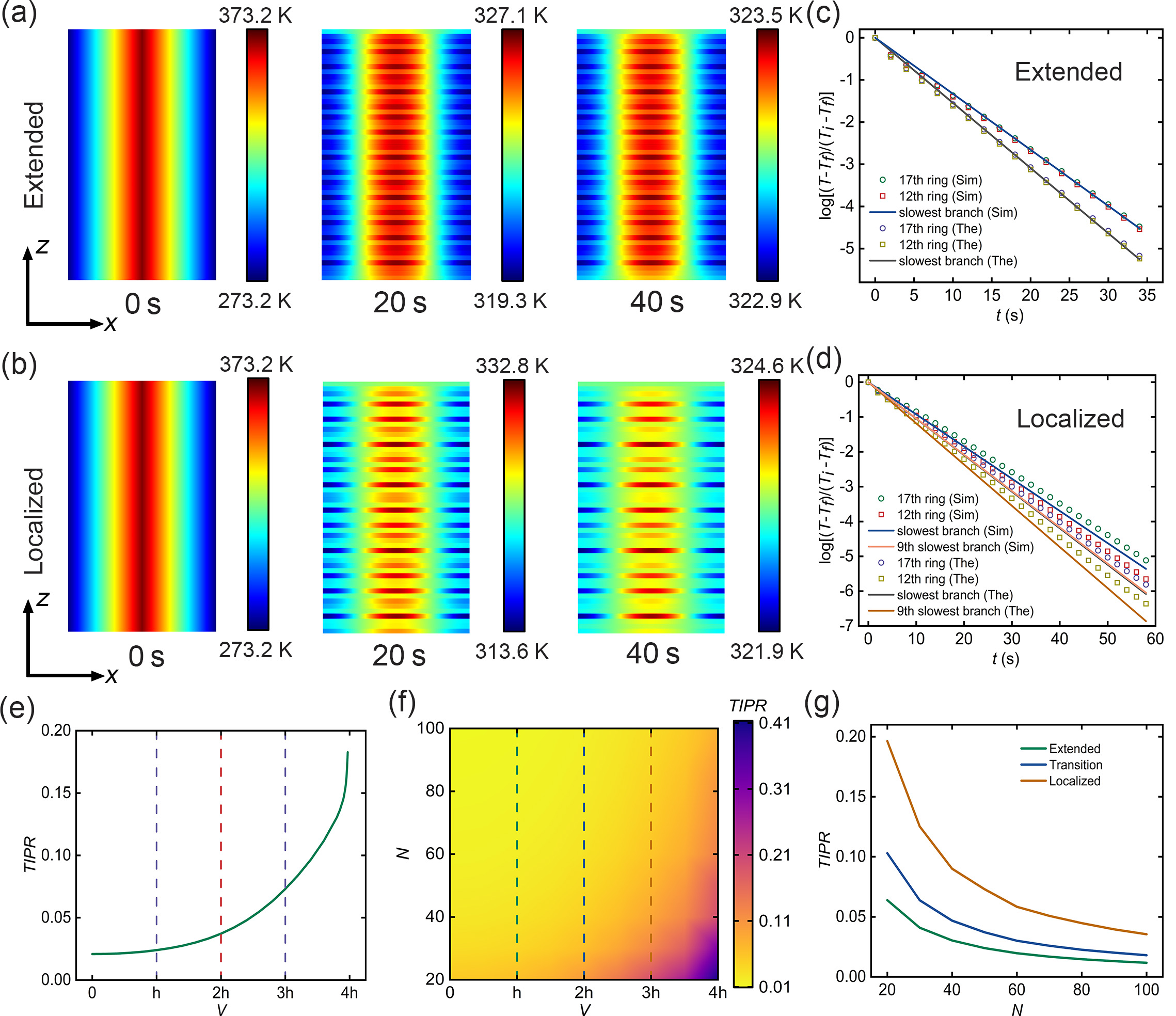}
\caption{Temperature field simulations of the extended-localized transition under the uniform excitation. (a) represents the extended state ($V=h$), while (b) depicts the localized state ($V=3h$). (c) The simulated and theoretical normalized maximum temperature evolutions of the 17th and 12th rings for the extended state. The blue and grey lines are the reference lines predicted by the simulated and theoretical decay rates of the slowest decaying branch. (d) The simulated and theoretical normalized maximum temperature evolutions of the 17th and 12th rings for the localized state. The blue and yellowish pink lines are the reference lines predicted by the simulated decay rates of the slowest and 9th slowest decaying branches. The grey and brown lines are the reference lines predicted by the theoretical decay rates of the slowest and 9th slowest decaying branches. The number of rings is $N=50$. (e) The $TIPR$ with different strengths of quasiperiodic potential. The temperature field used to calculated the $TIPR$ is the final state (see Appendix~\ref{app:AppB}). (f) Phase diagram of $TIPR$ with different numbers of rings $N$ and strengths of quasiperiodic potential $V$. (g) The $TIPR$ with different numbers of rings for the extended state, the transition point, and the localized state, which are marked as dashed lines in Fig.~\ref{Fig3}(f). The highest and lowest temperatures in the temperature field simulations are set as $T_{h}=373.2$~K and $T_{l}=273.2$~K. In the $y$-axis of Figs.~\ref{Fig3}(c,d), $T_{f}=(T_{h}+T_{l})/2=323.2$~K and $T_{i}=T_{h}=373.2$~K. The parameters are the same as in Fig.~\ref{Fig2}.}
\label{Fig3}
\end{figure}

Next we perform temperature field simulations to investigate the extended-localized transition. Firstly, we choose the uniform excitation as the initial condition (see Appendix~\ref{app:AppB}). For the diffusive extended state at $V=h$, the temperature field is evenly distributed and concentrates into the center, as depicted in Fig.~\ref{Fig3}(a). While for the diffusive localized state at $V=3h$ in Fig.~\ref{Fig3}(b), the temperature distribution becomes highly localized with multiple distinct localization centers. This multiplicity comes from the closeness of decay rates among several slow decaying branches, and is distinct in diffusion systems without the wave counterpart. Consequently, the temperature field simulations of the diffusive AAH model clearly demonstrate the presence of an extended-localized transition.

We extract the temperature evolutions to delve deeper into the quantitative analysis of the thermal behaviour within the extended-localized transition. Firstly, we extract the maximum temperature of the 17th ring. This is the localization ring of the slowest decaying branch for the localized state. For a comparative analysis, we also extract the maximum temperature of the 12th ring, which is the localization ring of the 9th slowest decaying branch for the localized state. The distribution of uniform excitation is close to the one of extended state. So for the diffusive extended state, the simulated (theoretical) maximum temperature evolutions of both rings correspond with each other and are predicted by the decay rate of the simulated (theoretical) slowest decaying branch, as shown in Fig.~\ref{Fig3}(c). The theoretical results solved by thermal coupling equations (see Appendix~\ref{app:AppC}) slightly deviate from the simulated ones due to the slight difference between the theoretical tight-binding model and the simulated structure. On the other hand, the uniform excitation can be approximatively considered as the superposition of all localized state distributions. So for the localized state, the temperature evolution of each ring is close to the reference line predicted by its corresponding localized branch, which is evident in Fig.~\ref{Fig3}(d). Besides, the temperature evolutions of the 17th and 12th rings decay slightly slower than the reference lines of the slowest and 9th slowest branches. The reason is that the maximum temperature decay of both rings is slowed down by their adjacent rings, which also have a high temperature region under the uniform initial excitation. Furthermore, the maximum temperature of the 17th ring decays slower than the one of the 12th ring because of the decay rate difference between the slowest and 9th slowest decaying branches. 

The temperature field alone can only tell us the degree of localization qualitatively. In order to quantitatively characterize the localization of temperature field, we should introduce the temperature $IPR$ ($TIPR$), which is the thermal analogue of $IPR$ in the condensed matter physics. The $TIPR$ is defined as 
\begin{equation}\label{6} 
TIPR=\frac{\sum_{j}(T_{j,\rm{max}}-T_{j,\rm{min}})^4}{\left(\sum_{j}{(T_{j,\rm{max}}-T_{j,\rm{min}})^2}\right)^2}
\end{equation}
where $T_{j,\rm{max}}$ and $T_{j,\rm{min}}$ are the maximum and minimum temperatures of the $j$-th ring. Similar to the $IPR$, the $TIPR$ approaches to zero for the extended state and becomes large for the localized state. The $TIPR$ for the diffusive AAH model ($N=50$) is shown in Fig.~\ref{Fig3}(e). With the increase of strength of quasiperiodic potential, the $TIPR$ also increases similar with $IPR$. However, the $TIPR$ for the localized state is evidently smaller than $IPR$ [see Fig.~\ref{Fig2}(a)]. The reason is that the decay rates of several branches are close to the one of the slowest decaying branch, so in the temperature field these branches will also emerge in addition to the slowest branch, which will reduce the $TIPR$. 

Next we discuss the size effect of temperature field simulations for diffusive AAH model according to the $TIPR$. Figure~\ref{Fig3}(f) presents the phase diagram of $TIPR$ with varying numbers of rings $N$ and strengths of quasiperiodic potential $V$. Besides, we extract the $TIPR$ with different $N$ for the extended state, transition point, and localized state, as demonstrated in Fig.~\ref{Fig3}(g). When $V$ keeps unchanged, the $TIPR$ exhibits a decreasing trend with the increasing $N$. This phenomenon can be attributed to more branches with a decay rate close to the slowest branch when $N$ increases, which will reduce the $TIPR$.

\section{\label{sec:local_excit}Temperature field simulations of diffusive AAH model under the local excitation} 

We proceed by altering the initial condition to local excitation (see Appendix~\ref{app:AppB}). As depicted in Fig.~\ref{Fig4}(a), the temperature field for the extended state is uniformly distributed amongst different localization centers. In contrast, for the localized state, as evident in Fig.~\ref{Fig4}(b), the temperature field remains confined within multiple localization centers. As before, we extract the maximum temperature evolutions of the 17th and 12th rings. For the extended state, the maximum temperature evolutions for both rings largely deviate from the reference lines predicted by the slowest decaying branch under the local excitation, which is demonstrated in Fig.~\ref{Fig4}(c). The reason is that the local excitation does not conform to the eigenstate distribution of the extended state, so the temperature evolution will not follow the decay rate of the slowest decaying branch. Nevertheless, for the localized state, as the eigenstate distributions align closer with the initial condition, the temperature evolutions for both rings deviate smaller from the reference lines anticipated by their respective branches [see in Fig.~\ref{Fig4}(d)] than the ones for the extended state. In contrast to the uniform excitation, the temperature evolution for both rings decays slightly faster than their respective branches because the initial temperature field of their adjacent rings is at the thermal equilibrium under the local excitation. With different initial conditions, the $TIPR$ with the local excitation [as depicted in Fig.~\ref{Fig4}(e)] is slightly larger than the one with the uniform excitation [refer to Fig.~\ref{Fig3}(e)]. The phase diagram of $TIPR$ with varying numbers of rings $N$ and strengths of quasiperiodic potential $V$ is shown in Fig.~\ref{Fig4}(f). We also extract the $TIPR$ with different $N$ for the extended state, transition point, and localized state, as demonstrated in Fig.~\ref{Fig4}(g). The results for the size effect under the local excitation are akin to the ones under the uniform excitation. Furthermore, we have demonstrated the temperature field simulations for the gradient and random initial excitations in Appendix~\ref{app:AppD}, which shows the universality of the extended-localized transition under more initial conditions. We have also performed the temperature field simulations under the case of $V=0$ in Appendix~\ref{app:AppE}. The temperature field for a periodic system shows a very uniform distribution under different excitations [see Fig.~\ref{Fig10}], which verifies that the quasi-disorder of parameters is the key to the multiple localization centers phenomena. 

\begin{figure}
\includegraphics[width=0.75\linewidth]{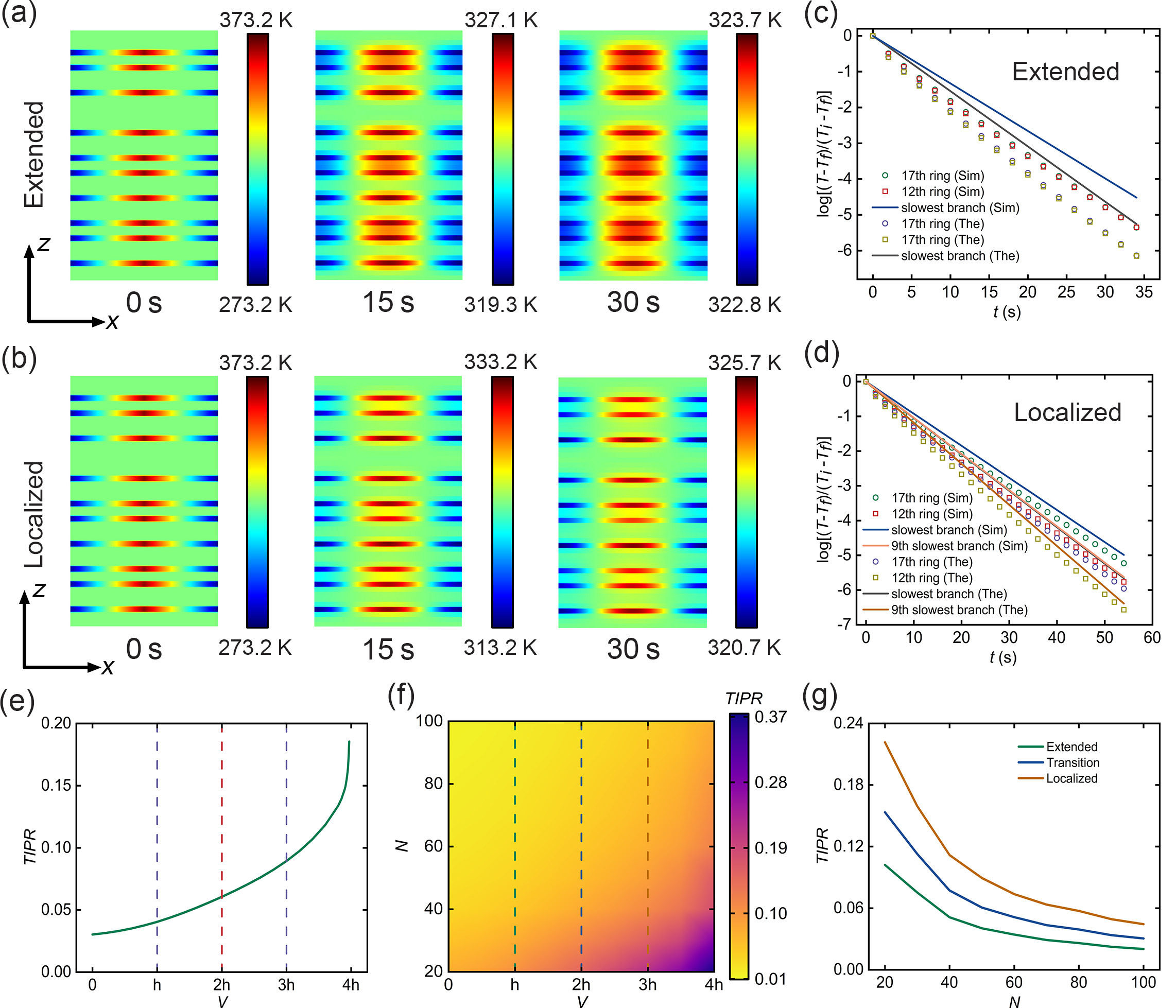}
\caption{Temperature field simulations of the extended-localized transition under the local excitation. (a) represents the extended state ($V=h$), while (b) depicts the localized state ($V=3h$). (c) The simulated and theoretical normalized maximum temperature evolutions of the 17th and 12th rings for the extended state. (d) The simulated and theoretical normalized maximum temperature evolutions of the 17th and 12th rings for the localized state. (e) The $TIPR$ with different strengths of quasiperiodic potential. (f) Phase diagram of $TIPR$ with different numbers of rings $N$ and strengths of quasiperiodic potential $V$. (g) The $TIPR$ with different numbers of rings for the extended state, the transition point, and the localized state, which are marked as dashed lines in Fig.~\ref{Fig4}(f). Other parameters are the same as in Fig.~\ref{Fig3}.}
\label{Fig4}
\end{figure}

\section{\label{sec:APT}Diffusive APT symmetric AAH model and its temperature field simulations}

\begin{figure}
\includegraphics[width=0.6\linewidth]{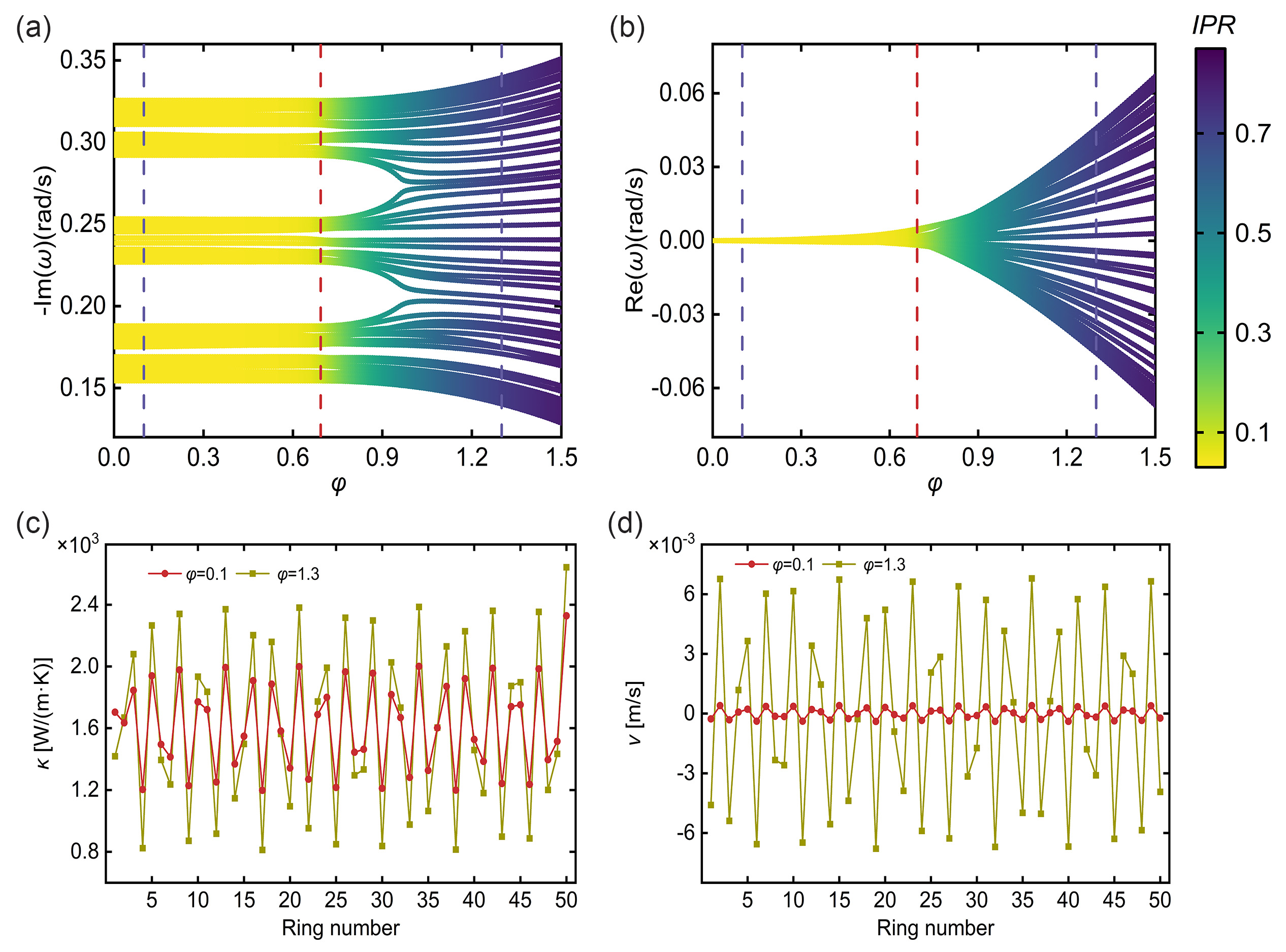}
\caption{Diffusive APT symmetric AAH model. (a) Decay rates and (b) eigenfrequencies with different imaginary phases. The colorbar indicates the $IPR$. The red dashed line marks the extended-localized (APT unbroken-broken) transition point $\varphi={\rm{log}}(2h/V)={\rm{log}(2)}{\approx}0.693$. The adjusted (c) thermal conductivities and (d) rotating velocities of each ring for the extended state ($\varphi=0.1$) and localized state ($\varphi=1.3$), which are marked as purple dashed lines in Figs.~\ref{Fig5}(a,b). Here we set $V=h$. Other parameters are the same as in Fig.~\ref{Fig2}.}
\label{Fig5}
\end{figure}

Next we introduce an imaginary phase to implement the diffusive APT symmetric AAH model. Similar to the derivation of the diffusive AAH Hamiltonian, the diffusive APT symmetric AAH Hamiltonian can be expressed as
\begin{equation}\label{7}
\hat{H}=i\sum_{j}\left[h\hat{c}_{j+1}^{\dagger}\hat{c}_{j}+h\hat{c}_{j}^{\dagger}\hat{c}_{j+1}-\left(V{\rm{cos}}(2\pi\alpha{j}+i\varphi)+Mh\right)\hat{c}_{j}^{\dagger}\hat{c}_{j}\right]
\end{equation}
where $\varphi$ represents the imaginary phase. The extended-localized transition occurs at the APT transition point $\varphi_{c}={\rm{log}}(2h/V)$~\cite{LonghiPRL19}. This exotic property has been numerically verified through the decay rates and eigenfrequencies of diffusive APT symmetric AAH model, as shown in Figs.~\ref{Fig5}(a,b). From the spectrum we find that the extended state is in the APT unbroken phase, while the localized state is in the APT broken phase.
 
Furthermore, the topological properties of the diffusive APT symmetric AAH model can be characterized by a winding number~\cite{LonghiPRL19}. By introducing a real phase $\phi$ into the onsite potential, the Hamiltonian can be expressed as 
\begin{equation}\label{8}
\hat{H}(\phi)=i\sum_{j}\left[h\hat{c}_{j+1}^{\dagger}\hat{c}_{j}+h\hat{c}_{j}^{\dagger}\hat{c}_{j+1}-\left(V{\rm{cos}}(2\pi\alpha{j}+\phi+i\varphi)+Mh\right)\hat{c}_{j}^{\dagger}\hat{c}_{j}\right]
\end{equation}
The winding number $\nu_{\phi}$ can be defined as
\begin{equation}\label{9}
\nu_{\phi}=\frac{i}{2{\pi}}\frac{1}{N}\int_{0}^{2\pi}d{\phi}{\partial}_{\phi}{\,}{\rm{log\,det}}\,\left[\hat{H}({\phi})-E_{B}\right]
\end{equation} 
where $E_{B}$ is the base energy. For the extended state, $\nu_{\phi}=0$. In contrast, for the localized state, $\nu_{\phi}=-1$ when $\varphi>0$ and $\nu_{\phi}=1$ when $\varphi<0$. 

For the diffusive APT symmetric AAH model with a complex onsite potential, the rotating terms of the rings should be introduced and adjusted, in addition to their thermal conductivities. By establishing a mapping between the diffusive APT symmetric AAH model and the ring chain's Hamiltonian, we obtain the following relations:
\begin{equation}\label{10}
\begin{aligned}
S_{1}+i{\beta}v_{1}&=-\left(V{\rm{cos}}(2\pi\alpha+i\varphi)+Mh\right), \\ 
S_{j}+i{\beta}v_{j}&=-\left(V{\rm{cos}}(2\pi{\alpha}j+i\varphi)+Mh\right), \\ 
S_{N}+i{\beta}v_{N}&=-\left(V{\rm{cos}}(2\pi{\alpha}N+i\varphi)+Mh\right),
\end{aligned}
\end{equation}
Here, $j=2,\cdots,N-1$. Note that the real part of the onsite potential corresponds to the onsite term of each ring, while the imaginary part corresponds to the rotation term. The solution of Eq.~\ref{10} provides the adjusted thermal conductivities and rotating velocities of the rings. To illustrate this point, we consider an extended state ($\varphi=0.1$) and a localized state ($\varphi=1.3$) as two examples. The distributions of the adjusted thermal conductivities and rotating velocities for these states are presented in Figs.~\ref{Fig5}(c,d). 

\begin{figure}
\includegraphics[width=0.75\linewidth]{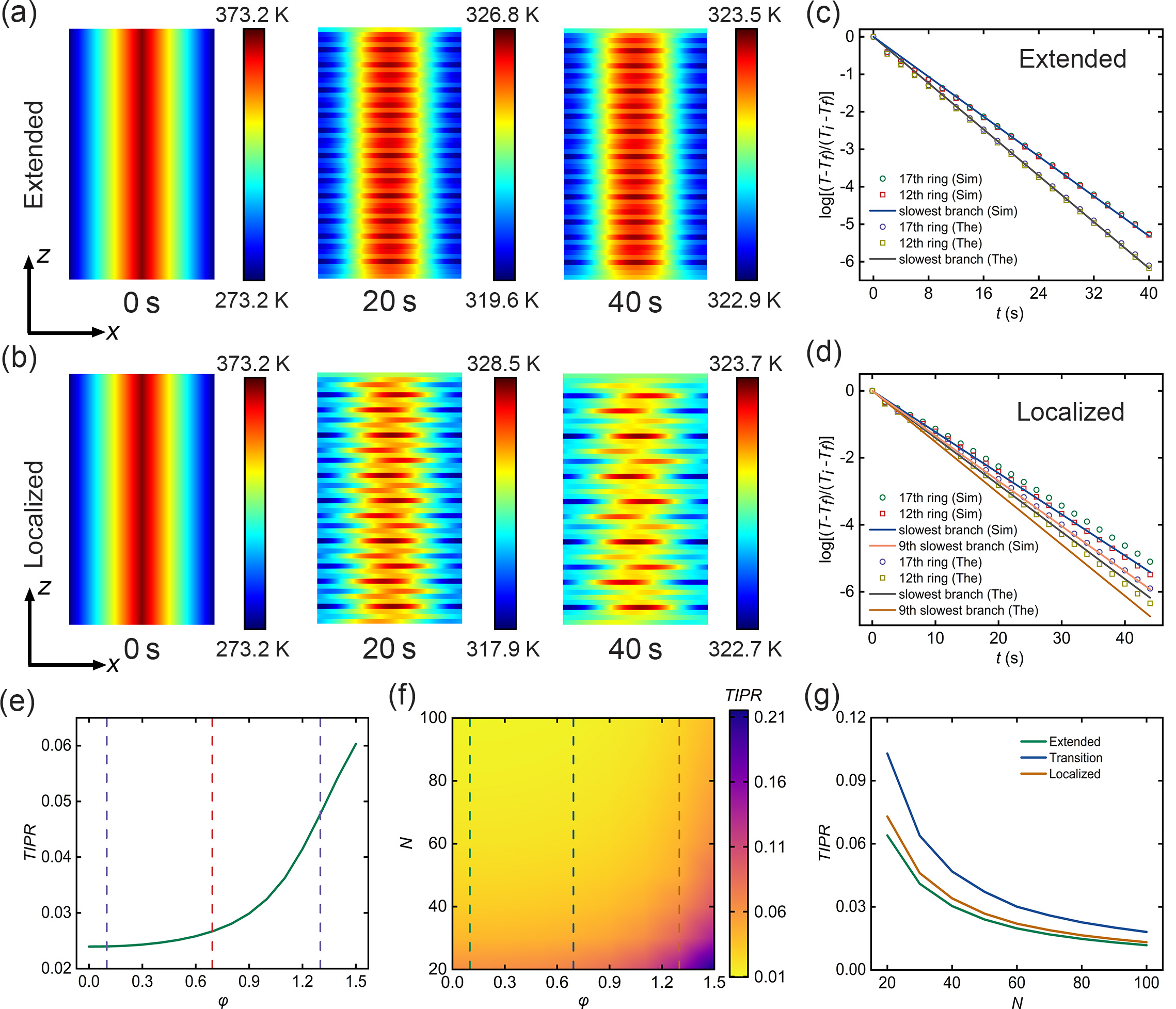}
\caption{Temperature field simulations for diffusive APT symmetric AAH model. (a) represents the extended state ($\varphi=0.1$), while (b) depicts the localized state ($\varphi=1.3$). (c) The simulated and theoretical normalized maximum temperature evolutions of the 17th and 12th rings for the extended state. (d) The simulated and theoretical normalized maximum temperature evolutions of the 17th and 12th rings for the localized state. (e) The $TIPR$ with different imaginary phases. (f) Phase diagram of $TIPR$ with different numbers of rings $N$ and imaginary phase $\varphi$. (g) The $TIPR$ with different numbers of rings for the extended state, the transition point, and the localized state, which are marked as dashed lines in Fig.~\ref{Fig6}(f). The parameters are the same as in Fig.~\ref{Fig3}.}
\label{Fig6}
\end{figure}

Next, we perform temperature field simulations for the diffusive APT symmetric AAH model under the uniform excitation. For the extended state ($\varphi=0.1$) [see in Fig.~\ref{Fig6}(a)], the temperature field demonstrates a uniform distribution and remains stationary. Conversely, for the localized state when $\varphi=1.3$, the temperature distribution distinctly reveals moving multiple localization centers, as depicted in Fig.~\ref{Fig6}(b). This observation suggests a simultaneous occurrence of an extended-localized transition [represented by an increasing $TIPR$ with $\varphi$ in Fig.~\ref{Fig6}(e)] and an APT unbroken-broken transition for the diffusive APT symmetric AAH model. Additionally, we extract the maximum temperature evolutions of the 17th and 12th rings. For the extended state, the evolutions of maximum temperatures for both rings align with the reference lines predicted by the slowest decaying branch, as shown in Fig.~\ref{Fig6}(c). Meanwhile, for the localized state [see Fig.~\ref{Fig6}(d)], the maximum temperature of the 17th ring decays slower than the one of the 12th ring because of the decay rate difference between the slowest and 9th slowest decaying branches. Besides, the temperature evolutions of both rings decay slightly slower than the reference lines predicted by the corresponding branches. The phase diagram of $TIPR$ with varying numbers of rings $N$ and imaginary phases $\varphi$ is shown in Fig.~\ref{Fig6}(f). We also extract the $TIPR$ with different $N$ for the extended state, transition point, and localized state, as demonstrated in Fig.~\ref{Fig6}(g). The results for the temperature evolution and size effect of diffusive APT symmetric AAH model are similar with the ones of diffusive AAH model.

\section{\label{sec:exper_sugg}Experimental suggestions}

Now we provide some experimental suggestions for implementing the diffusive AAH model. The experimental setup proposed in Ref.~\cite{XuNP22} can be utilized to construct the coupled ring chain structure. However, it is crucial to ensure that the parameters of rings should conform to the results of our theoretical calculations. We suggest using well-designed composite materials based on effective medium theory to meet this requirement. For instance, adjusting the doping rates of materials with high contrast, such as copper and Polydimethylsiloxane, can help achieve the necessary material parameters. Furthermore, advection can provide a reasonable degree of freedom for flexibly tuning effective thermal conductivity~\cite{LiNM19, XuNC20}.

\section{\label{sec:potent_appli}Potential application: double-trace distributed generator}

\begin{figure}
\includegraphics[width=0.5\linewidth]{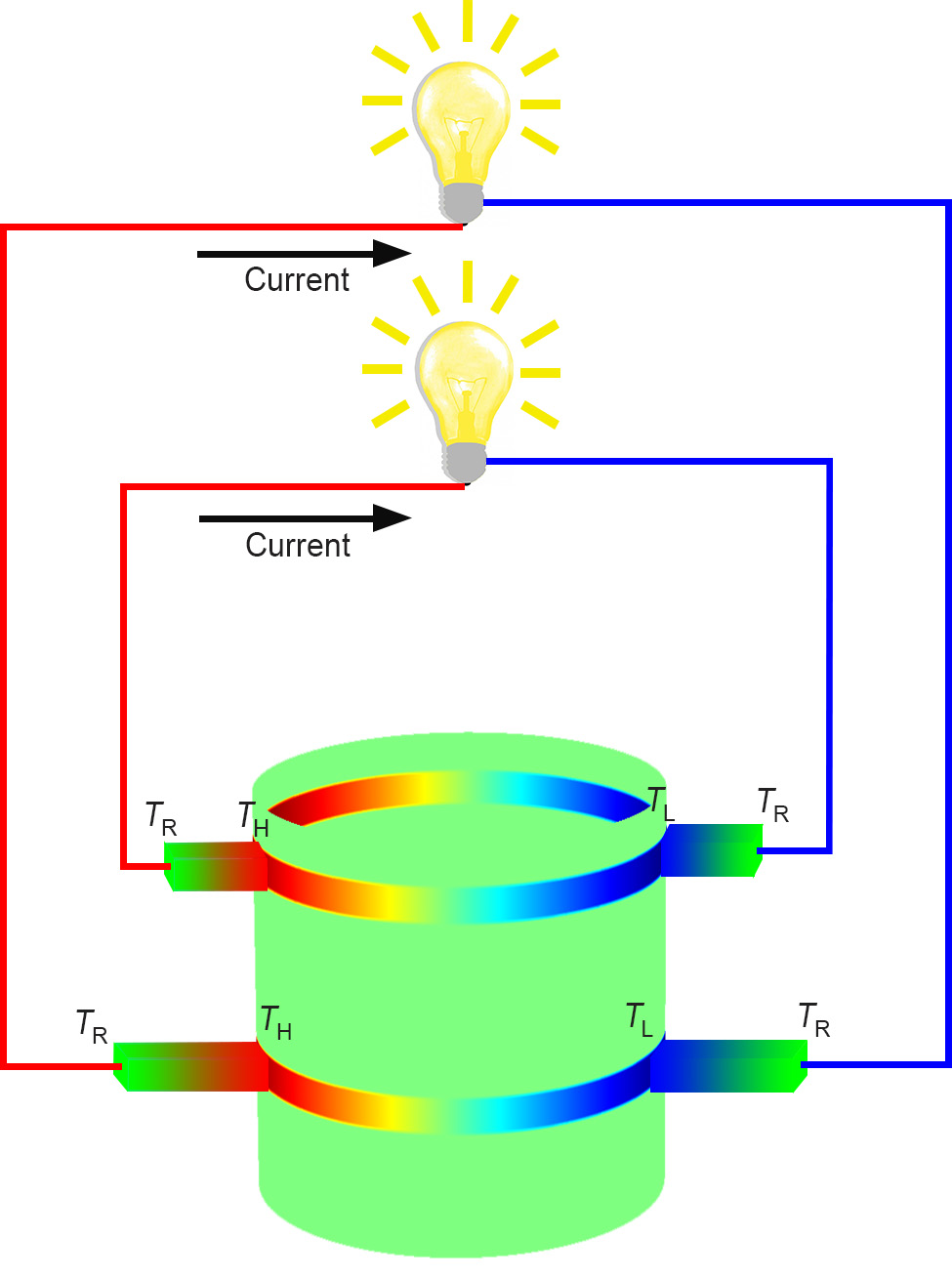}
\caption{Double-trace distributed generator. The cuboid indicates the thermoelectric material. The temperature at the hot source, the temperature at the cold source, and the room temperature are denoted by $T_{\rm{H}}$, $T_{\rm{L}}$, and $T_{\rm{R}}$, respectively.}
\label{Fig7}
\end{figure}

Then we propose a potential application of double localization centers (multiple localization centers by extension naturally), referred to as double-trace distributed generator. In a thermal system, high temperature waste heat and low temperature waste heat are often present. By connecting the two positions of a coupled ring chain structure to high and low temperature waste heat sources, and incorporating thermoelectric materials, external electricity can be generated. When the thermal system ceases to function, the high and low temperature heat sources disappear, causing the temperature field of the structure to evolve and leading to the emergence of the double localization centers phenomenon. As shown in Fig.~\ref{Fig7}, we can choose the positions of thermoelectric materials as where the double localization centers appear. By doing so, the power and duration of the thermoelectric materials can be significantly enhanced compared to a structure without adjusted parameters, and two objects can be powered simultaneously. Our device has a straightforward structure and flexible distributed power generation capabilities when compared to traditional electric generators.

\section{\label{sec:conclusion}Conclusion}

In this study, we have successfully demonstrated the extended-localized transition within diffusion systems. This transition is realized through a diffusive quasicrystal constructed using a coupled ring chain structure. By carefully adjusting the thermal conductivities of rings, we have established the diffusive AAH model successfully. Through temperature field simulations, we have clearly confirmed the extended-localized transition for the diffusive AAH model based on this ring-shaped chain. The temperature evolution and size effect of this transition are further studied quantitatively. Meanwhile, we utilize the initial condition of local excitation to investigate the temperature field of both the extended and localized states. Furthermore, we introduce the non-Hermitian physics into diffusive AAH model to investigate the extended-localized transition. This diffusive APT symmetric AAH model shows a moving multiple localization centers phenomenon in the temperature field simulation. At last, we provide some experimental suggestions and propose a potential application. We anticipate that our findings will stimulate further researches into quasi-disordered and disordered phases in diffusion systems~\cite{TianAPL24, LiuCPL23}. For instance, the study of the integer quantum Hall insulator in diffusion systems can be carried out due to its mapping to the diffusive AAH model~\cite{LangPRL12, KrausPRL12}. The coupled ring chain structure can be used to realize various topological states in diffusion systems~\cite{CaoPER22, ZhengPER23, LiPER23}. Besides, these exotic phases of matter can help in the design of innovative thermal materials for efficient and robust heat manipulation~\cite{ZhouNC23}.

\appendix

\section{\label{app:AppA}Lyapunov exponent of diffusive AAH model}

\begin{figure}
\includegraphics[width=0.5\linewidth]{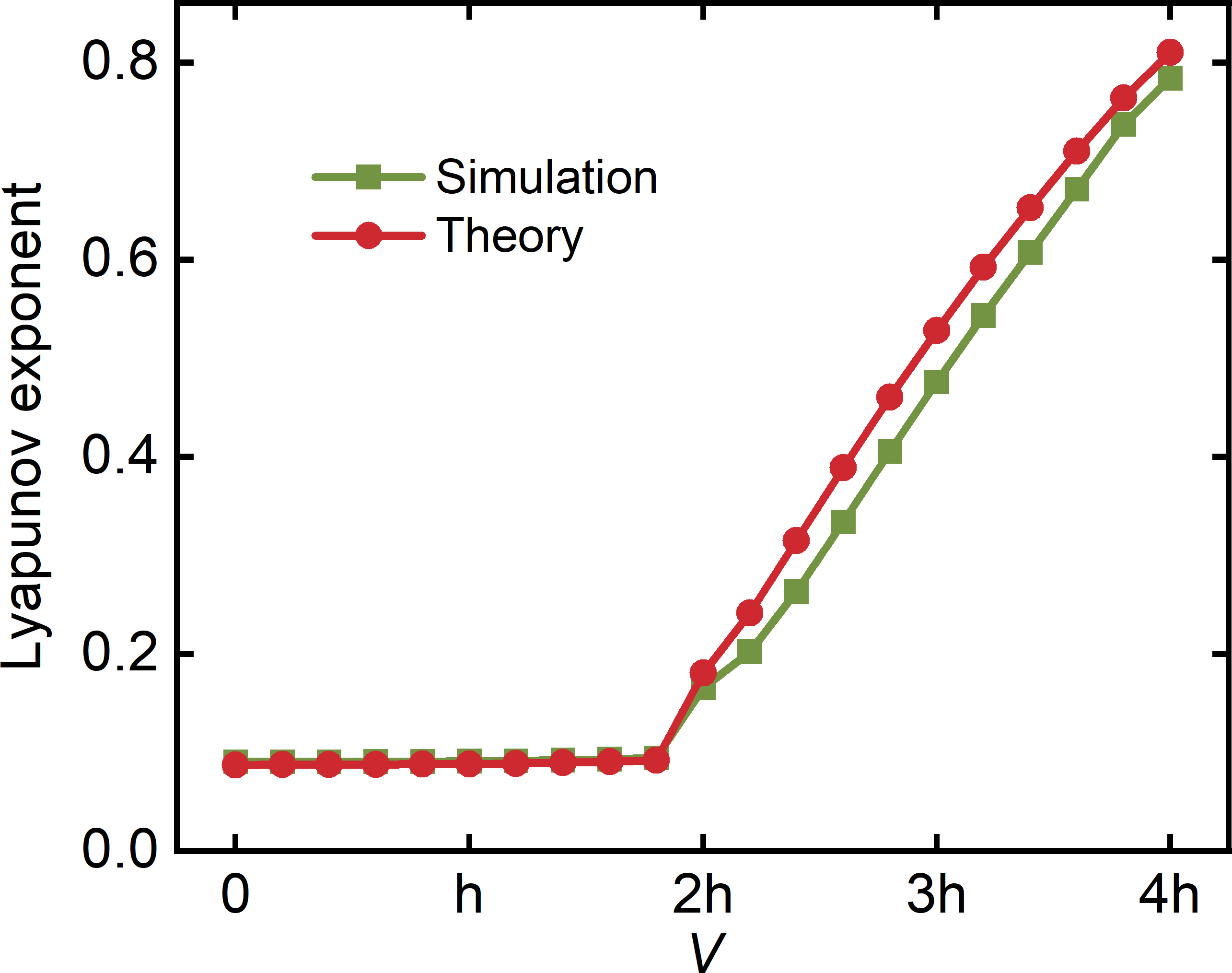}
\caption{Theoretical and simulated Lyapunov exponent of the slowest decaying branch. The parameters are the same as in Fig.~\ref{Fig2}.}
\label{Fig8}
\end{figure}

Here we want to extract the Lyapunov exponent of diffusive AAH model. For the diffusive AAH model discussed in the main text, the extended-localized transition occurs at $V=2h$ due to the self-duality~\cite{Aubry80}. The model is in the extended phase when $V<2h$ while in the localized phase if $V>2h$. Furthermore, the localized eigenstate has an exponentially localized form with $|{\psi}|=|{\psi}|_{\rm{max}}e^{-{\eta}|i-i_{0}|}$, where $i_{0}$ is the ring index of localization center and ${\eta}$ is the Lyapunov exponent~\cite{JiangPRB19}. The Lyapunov exponent, similar with the $IPR$, also characterizes the degree of localization. However, it is difficult to extract the Lyapunov exponent from the temperature field. The reason is that several slow decaying branches will emerge in the temperature field due to their close decay rates, as can be seen from the Fig.~\ref{Fig2}(a). However, we can extract the Lyapunov exponent from the theoretical and simulated eigenstate distributions by exponential fitting. The results for the Lyapunov exponent of the slowest decaying branch are shown in the Fig.~\ref{Fig8}, where a clear extended-localized phase transition can be found. 

\section{\label{app:AppB}Temperature field simulations under the uniform and local excitations}

We perform the temperature field simulations by COMSOL Multiphysics. The ambient temperature is set to be $293.2$~K. We stop the simulation when the difference between maximum and minimum temperatures of the structure drops below $0.5$~K. We call the temperature field under this condition as the final state. For the uniform excitation, we set the initial temperature at the middle of channel $T_h$ as the highest, and one at the ends (periodic boundary condition) $T_l$ as the lowest. The initial temperature field between these two positions is linearly distributed. In the case of local excitation, we stimulate $N/5$ rings, which are the localization rings of $N/5$ slow decaying branches for the localized state. The temperatures of the remaining rings are set at the thermal equilibrium $T_{f}=(T_{h}+T_{l})/2$. For the diffusive AAH model ($N=50$) discussed in the main text, these ten rings are ranked as follows: 17th (slowest), 38th (2nd slowest), 4th (3rd slowest), 30th (4th slowest), 25th (5th slowest), 9th (6th slowest), 46th (7th slowest), 43rd (8th slowest), 12th (9th slowest), and 22nd (10th slowest). 

\section{\label{app:AppC}The theoretical solution for thermal coupling equations}

The detailed theoretical solution for thermal coupling equations involves spatially discretizing the ring structure and then solving the partial differential equations. As shown in the main text, the temperature field $T_{j}(x,t)$ of the $j$-th ring is expressed in the thermal coupling equation: 
\begin{align}\label{11}
&\frac{\partial T_{j}(x,t)}{\partial t}=\frac{\kappa_j}{\rho_j C_j}\frac{\partial^2 T_{j}(x,t)}{\partial x^2}+v_j\frac{\partial T_{j}(x,t)}{\partial x}\nonumber \\
&+h_{j-1,j}\left[T_{j-1}(x,t)-T_{j}(x,t)\right]+h_{j,j}\left[T_{j+1}(x,t)-T_{j}(x,t)\right],
\end{align}  
Subsequently, we discretize each ring into $M$ segments, and designating the temperature field for each segment of the $j$-th ring as $T_{j,k}(t)$, where $j=1,\cdots,N$ and $k=1,\cdots,M$. Additionally, due to the circular structure, the condition $T_{j,1}(t)=T_{j,M}(t)$ must be imposed. Hence, the discretized thermal coupling equation of the $j$-th ring can be expressed as
\begin{align}\label{12}
&\frac{\partial T_{j,k}(t)}{\partial t}=\frac{\kappa_j}{\rho_j C_j}\frac{T_{j,k+1}(t)+T_{j,k-1}(t)-2T_{j,k}(t)}{a^2}\nonumber \\
&+v_j\frac{T_{j,k+1}(t)-T_{j,k-1}(t)}{2a}+h_{j-1,j}\left[T_{j-1,k}(t)-T_{j,k}(t)\right]\nonumber \\
&+h_{j,j}\left[T_{j+1,k}(t)-T_{j,k}(t)\right],
\end{align}  
where $a=2{\pi}R/M$. Next, we substitute the parameters of the diffusive AAH model and the initial conditions (uniform/local excitation) into the discretized thermal coupling equations. Then we solve these equations numerically. After extracting the maximum temperature ${\rm{max}}\left[T_{j,1}(t),\cdots,T_{j,M}(t)\right]$ of the $j$-th ring, we obtain the theoretical results depicted in Figs.~\ref{Fig3}(c,d) and Figs.~\ref{Fig4}(c,d) within the main text.

\section{\label{app:AppD}Temperature field simulations of diffusive AAH model under the gradient and random excitations}

\begin{figure}
\includegraphics[width=\linewidth]{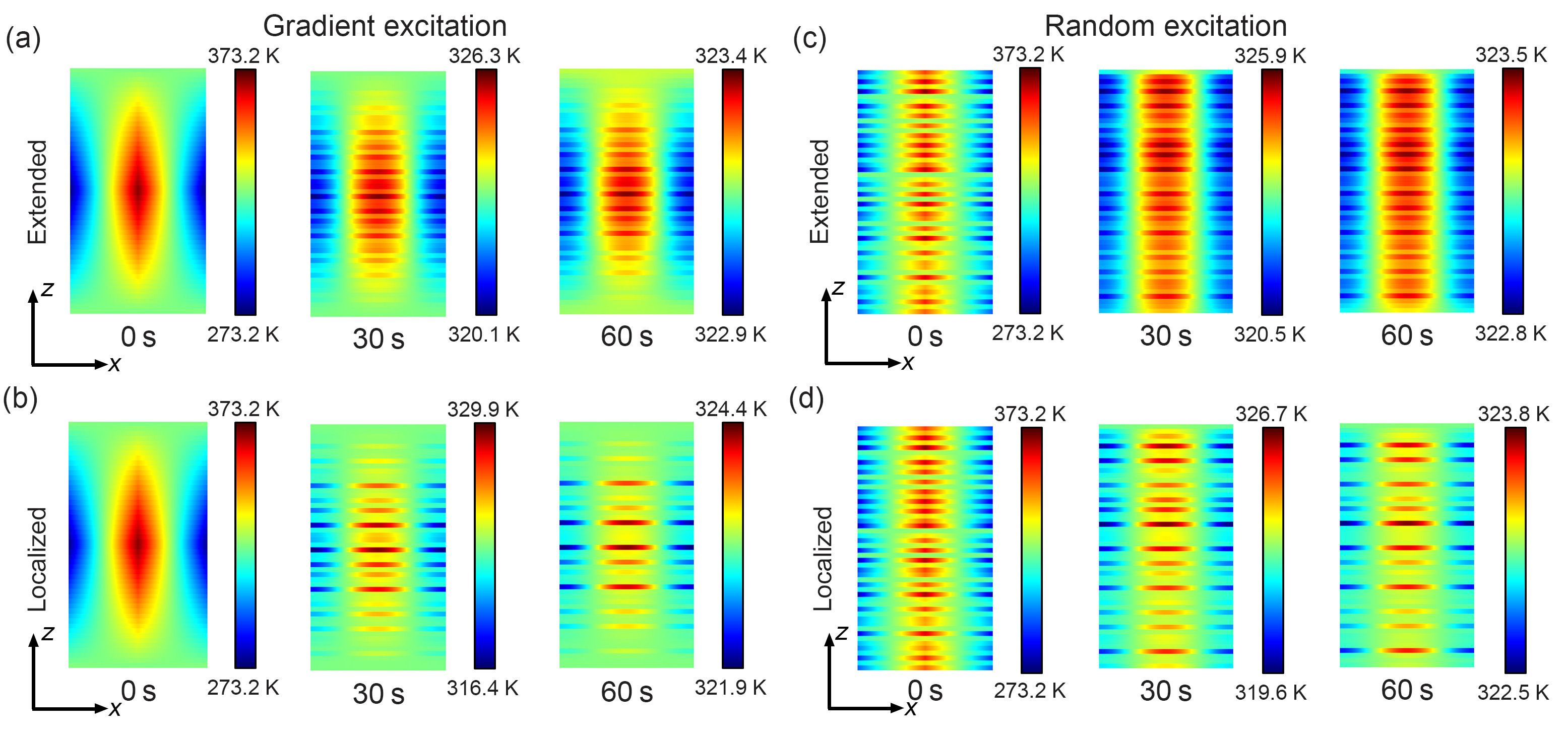}
\caption{Temperature field simulations under the gradient and random excitation. (a) represents the extended state ($V=h$) under the gradient excitation, while (b) depicts the localized state ($V=3h$) under the gradient excitation. (c) represents the extended state ($V=h$) under the random excitation, while (d) depicts the localized state ($V=3h$) under the random excitation. The parameters are the same as in Fig.~\ref{Fig3}.}
\label{Fig9}
\end{figure}

In this section, we would perform the temperature field simulations under other initial excitations except for the uniform and local excitations. Firstly, we set the gradient condition to excite the temperature field. For the gradient excitation, the temperature gradient for the central two rings has a maximum value and decreases in a gradient for other rings, while the initial temperature field for the boundary rings reaches the thermal equilibrium [see Figs.~\ref{Fig9}(a,b)]. The temperature field for the extended state concentrates to the center and has a similar distribution with the initial condition [see Fig.~\ref{Fig9}(a)]. However, the temperature field for the localized state exhibits a multiple localization centers phenomenon [see Fig.~\ref{Fig9}(b)]. Secondly, the random excitation is adopted to investigate the temperature field behaviour, which means that the initial temperature gradient for each ring is randomly imposed [see Figs.~\ref{Fig9}(c,d)]. Similar with other excitations, the temperature field for the extended state is uniform [see Fig.~\ref{Fig9}(c)] while the one for the localized state demonstrates several localization centers [see Fig.~\ref{Fig9}(d)]. Both gradient and random excitations have demonstrated an obvious extended-localized transition in the temperature field simulation.

\section{\label{app:AppE}Temperature field simulations without a quasiperiodic onsite potential}

\begin{figure}
\includegraphics[width=\linewidth]{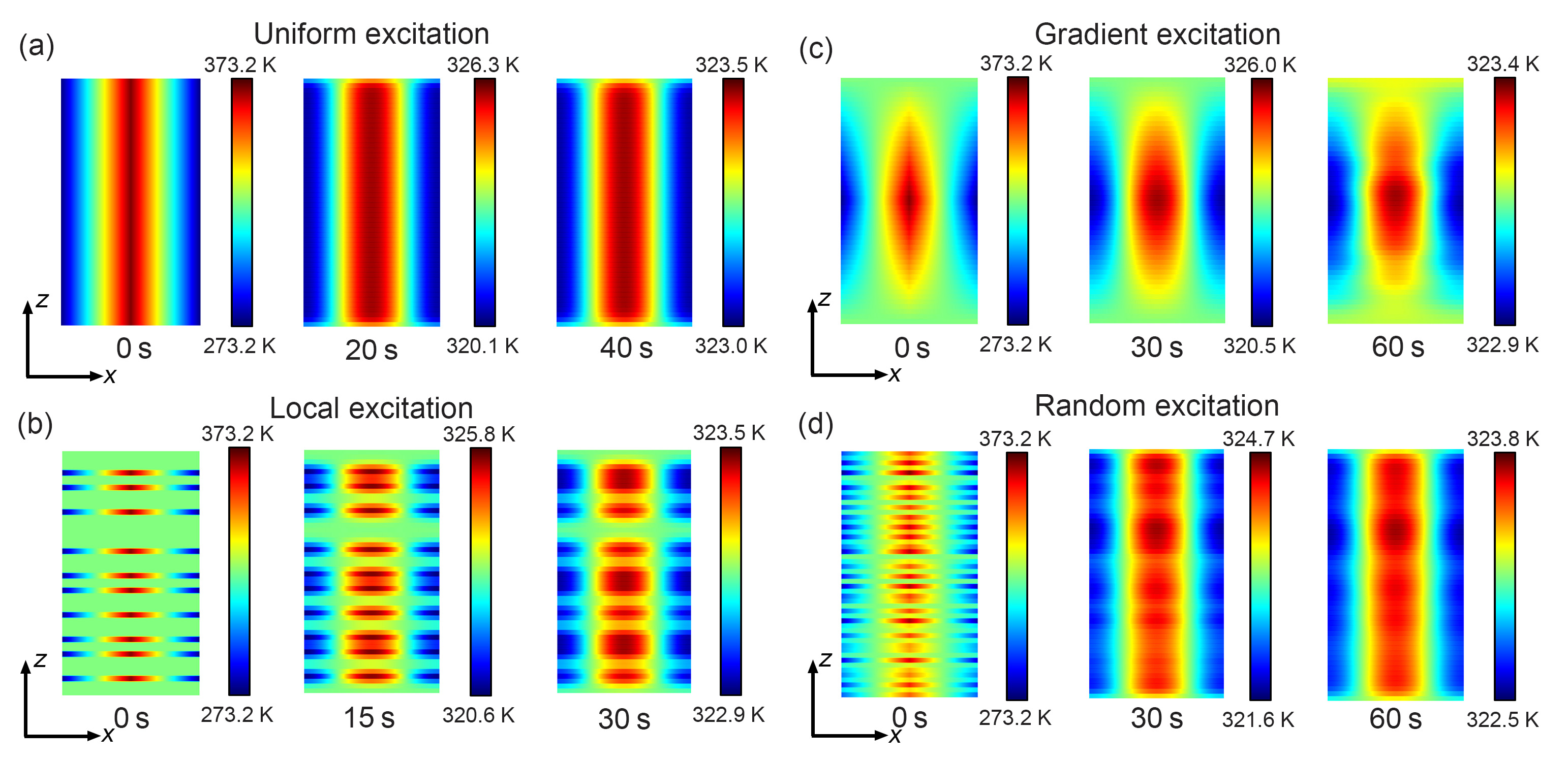}
\caption{Temperature field simulations without the quasiperiodicity. (a) shows the result under the uniform excitation. (b) shows the result under the local excitation. (c) shows the result under the gradient excitation. (d) shows the result under the random excitation. The parameters are the same as in Fig.~\ref{Fig3}.}
\label{Fig10}
\end{figure}

In this section, we would perform the temperature field simulations of diffusive AAH model with a zero onsite potential ($V=0$). As shown in Fig.~\ref{Fig10}, we have studied this issue under four initial conditions: uniform, local, gradient, and random excitations. For the uniform and random excitations, the temperature field demonstrates a fully uniform distribution [see Figs.~\ref{Fig10}(a,d)]. For the local excitation, the temperature field for $V=0$ [see Fig.~\ref{Fig10}(b)] is much more uniformly distributed amongst different localization centers than the ones for $V=h$ and $V=3h$ [see Figs.~\ref{Fig4}(a,b)]. For the gradient excitation, the temperature field concentrates to the center and has a similar distribution with the initial condition [see Fig.~\ref{Fig10}(c)]. The temperature fields for $V=0$ show a totally different distribution with the ones for $V=3h$ in the localized phase. So the multiple localization centers phenomenon actually stems from the quasi-disorder of the system.

\section*{Acknowledgements}
We are extremely grateful to Professor Ruibao Tao for his incredibly insightful discussions and guidance, and we also thank Dr. Xinchen Zhou for his valuable discussions. J. H. acknowledges the financial support provided by the National Natural Science Foundation of China under Grants No. 12035004 and No. 12320101004, and the Innovation Program of the Shanghai Municipal Education Commission under Grant No. 2023ZKZD06. Y. L. acknowledges the support by the National Natural Science Foundation of China under Grants Nos. 92163123 and 52250191, and Zhejiang Provincial Natural Science Foundation of China under Grant No. LZ24A050002.

\end{document}